\documentclass[twocolumn,letter]{IEEEtran}
\usepackage[cmex10]{amsmath}
\usepackage{siunitx}
\usepackage{cite}
\usepackage{psfrag}
\usepackage[utf8]{inputenc}
\usepackage[T1]{fontenc}
\usepackage{amsmath,amsfonts,amsbsy,amssymb}
\usepackage{mathabx}
\usepackage{mathrsfs}
\usepackage[nolist]{acronym}
\usepackage{tabularx}
\usepackage{amssymb}
\usepackage{amsmath}
\usepackage{graphicx}
\usepackage{cite}
\usepackage{multirow}
\usepackage{wasysym}
\usepackage{multirow}
\usepackage{float}
\usepackage{xcolor}
\usepackage{subcaption}
\usepackage{algorithm}
\usepackage{algorithmic}
\usepackage{xcolor}
\usepackage{suffix}
\usepackage[font=footnotesize]{caption}
\title{Virtual Microgrid Management via Software-defined Energy Network for Electricity Sharing}
\author{Pedro H. J. Nardelli,~\IEEEmembership{Senior Member,~IEEE}, Hafiz Majid Hussain,~\IEEEmembership{Student Member,~IEEE,}, Arun Narayanan,~\IEEEmembership{Member,~IEEE}, and Yongheng Yang, ~\IEEEmembership{Senior Member,~IEEE}
\thanks{PHJN, HMH and AN are with LUT University, Finland. YY is with Department of Electrical Engineering, Zhejiang University. China. This paper is partly supported by Academy of Finland via: (a) ee-IoT project n.319009, (b) FIREMAN consortium CHIST-ERA/n.326270 (CHIST-ERA-17-BDSI-003), and (c) EnergyNet Research Fellowship n.321265/n.328869. The authors would like to thank Chris Giotitsas, Vasilis Kostakis and Simon Pirani for the discussions about the future of energy systems that partly motivated this work.}%
}

\begin{document}

\maketitle
%
\begin{abstract}
Digitalization has led to radical changes in the distribution of goods across various sectors.
The tendency is to move from traditional buyer--seller markets to subscription-based on-demand ``smart'' matching platforms enabled by pervasive information and communication technologies (ICTs).
The driving force behind this lies in the fact that assets, which were scarce in the past, are readily abundant, approaching a regime of zero marginal costs.
This is also becoming a reality in electrified energy systems due to the substantial growth of distributed renewable energy sources such as solar and wind; the increasing number of small-scale storage units such as batteries and heat pumps; and the availability of flexible loads that enable demand-side management.
In this context, this article proposes a system architecture based on a logical (cyber) association of spatially distributed (physical) elements as an approach to build a \textit{virtual microgrid} operated as a \textit{software-defined energy network} (SDEN) that is enabled by \textit{packetized energy management}.
The proposed cyber-physical system presumes that electrical energy is shared among its members and that the energy sharing is enabled in the cyber domain by handshakes inspired by resource allocation methods utilized in computer networks, wireless communications, and peer-to-peer Internet applications (e.g., BitTorrent). 
The proposal has twofold benefits: (i) reducing the complexity of current market-based solutions by removing unnecessary and costly mediations and (ii) guaranteeing energy access to all virtual microgrid members according to their individual needs. 
This article concludes that the proposed solution generally complies with the existing regulations but has highly disruptive potential to organize a dominantly electrified energy system in the mid- to long-term, being a technical counterpart to the recently developed social-oriented microgrid proposals.

\end{abstract}
 \begin{IEEEkeywords}
 packetized energy management, cyber-physical systems, electricity grid
 \end{IEEEkeywords}

\section{Introduction}
The astonishing development of information and communication technologies (ICTs) has brought widespread changes not only in the way people live and interact but also in how society and the economy are organized.
Digital technologies have led to solutions in different domains where the price of producing a new unit of a specific good tends to zero \cite{rifkin2014zero}.
This is the so-called zero marginal cost regime, where goods are abundant, and therefore, markets---which presuppose scarcity---do not function in the way their designers have expected.
For example, digital media such as downloadable music and e-books have nearly zero marginal costs\footnote{There is always some small storage cost for digital media, which is usually negligible in comparison to traditional production.} since the cost of producing one more unit of a copy of a given good is virtually zero, in contrast to a compact disc (CD) or a printed book where there is a non-zero production cost \cite{goldfarb2019digital}.
As a result, new business models like subscription-based access (e.g., Netflix and Spotify) and demand-based matching (e.g., Uber and Airbnb) are widely diffused now.
Such markets based on ICT-enabled platforms are not the only possibility.
An alternative governance model---frequently neglected by mainstream research---can be established based on a \textit{shared} pool of resources forming a \textbf{\textit{commons} }\cite{ostrom1990governing}.
The importance of this approach has been demonstrated by the awarding of the Nobel prize of Economics to Elinor Ostrom in 2009 for her ``analysis of economic governance, especially the commons.'' 

Since the late 20th century, the electricity sector has been moving toward a liberalized open market model. 
In this free market-based design, public utilities are unbundled to create liberalized markets comprising consumers, energy retailers, distribution network operators, transmission system operators, and large generation plants.
Today, the upsurge in the number of aggregators and other ancillary energy service providers, who heavily employ ICTs, are signaling some changes to the electricity market.
Changes are also precipitated by the fact that the electric grid is moving toward a scenario where a large fraction of the electricity costs is capital costs, and marginal operating costs are nearly zero \cite{lo2019electricity}. %
This development may well disrupt the current market structure or even make electricity markets obsolete, especially in the case of small-scale customers connected to the distribution grid, such as households or offices.

In Europe, electricity markets are evolving to accommodate new elements and zero marginal cost production (which have created negative prices during specific periods) \cite{immonen2020consumer}.
The ``Winter Package'' \cite{lavrijssen2017radical,ringel2018governance} is a piece of concrete evidence, where new regulations are being designed to support a large number of entities whose main role is to manage the distributed energy resources (DERs).
Remarkably, their management approaches strongly rely on ICTs creating a new series of cyber solutions like Virtual Power Plants, Virtual Batteries, and Virtual Microgrids (e.g., \cite{naval2020virtual, espinosa2020packetized, anoh2019energy,klein2019novel}), which have already been incorporated in the system operation.

Most of these solutions implicitly assume that energy is a scarce resource so that markets would be a natural solution to match supply and demand based on price signals \cite{mahmud2020internet}.
Additionally, their designs assume that an individual management strategy of DERs cannot affect the market price (i.e., individuals are price-takers), which is an unsound assumption if the number of elements following such a management algorithm is large (e.g., 50\% of households). 
For example, due to the current structure of European markets composed of day-ahead and balancing markets, the use of the so-called ``real-time price'' to take operational decisions, such as storing energy or postponing the operation of a flexible load, can be problematic due to the collective effects of individual optimizing decisions \cite{nardelli2018smart,kuhnlenz2018implementing}.
Other different approaches, such as peer-to-peer exchanges as in \cite{klein2019novel,zhou2018evaluation,klein2020pragmatic}, have also appeared as potential promising candidates for future governance models; here, \textit{prosumers} locally trade their production in local markets, also employing the most recent ICTs as distributed ledgers (blockchain) \cite{long2018peer}.

Given this background, this article sheds light on and puts forth a way to accomplish an alternative vision that assumes a \textit{commons-based governance model} \cite{giotitsas2015peer,giotitsas2020private} to build an Energy Internet based on packetized energy management \cite{nardelli2019energy,hussain2020energy}.
This vision also brings the advantages of decentralized optimization \cite{kristov2016tale,caballero2019social} but with a commons-based governance model.
Next, we will describe a software-defined energy network (SDEN) as a method to create a commons-based cyber-physical system where electricity is shared among all the virtual microgrid members.

\section{Packetized Energy Management and Software-defined Energy Network}
%
%
Although power grids have been traditionally managed to ensure that electrical supply follows electrical demand, the steady growth of distributed production units, which are highly dependent on intermittent weather conditions, has driven the development of new techniques that control \textit{flexible loads} to make the demand follow the availability of supply.
This is the already well-known \textit{demand-side management} (DSM) \cite{palensky2011demand,alizadeh2012demand,meyabadi2017review}.
DSM is a broad term that is associated with different timescales \cite{nardelli2019energy},\cite{palensky2011demand}, from primary control at (sub-)second-level to load/storage scheduling plans with time horizons of hours, days, and even months (if long-term storage is considered).
Furthermore, DSM is also related to different types of consumers, offering possibilities to, for instance, large-scale industries, commercial buildings, and households.

\begin{figure}[t]
\centering
  \includegraphics[width=\linewidth]{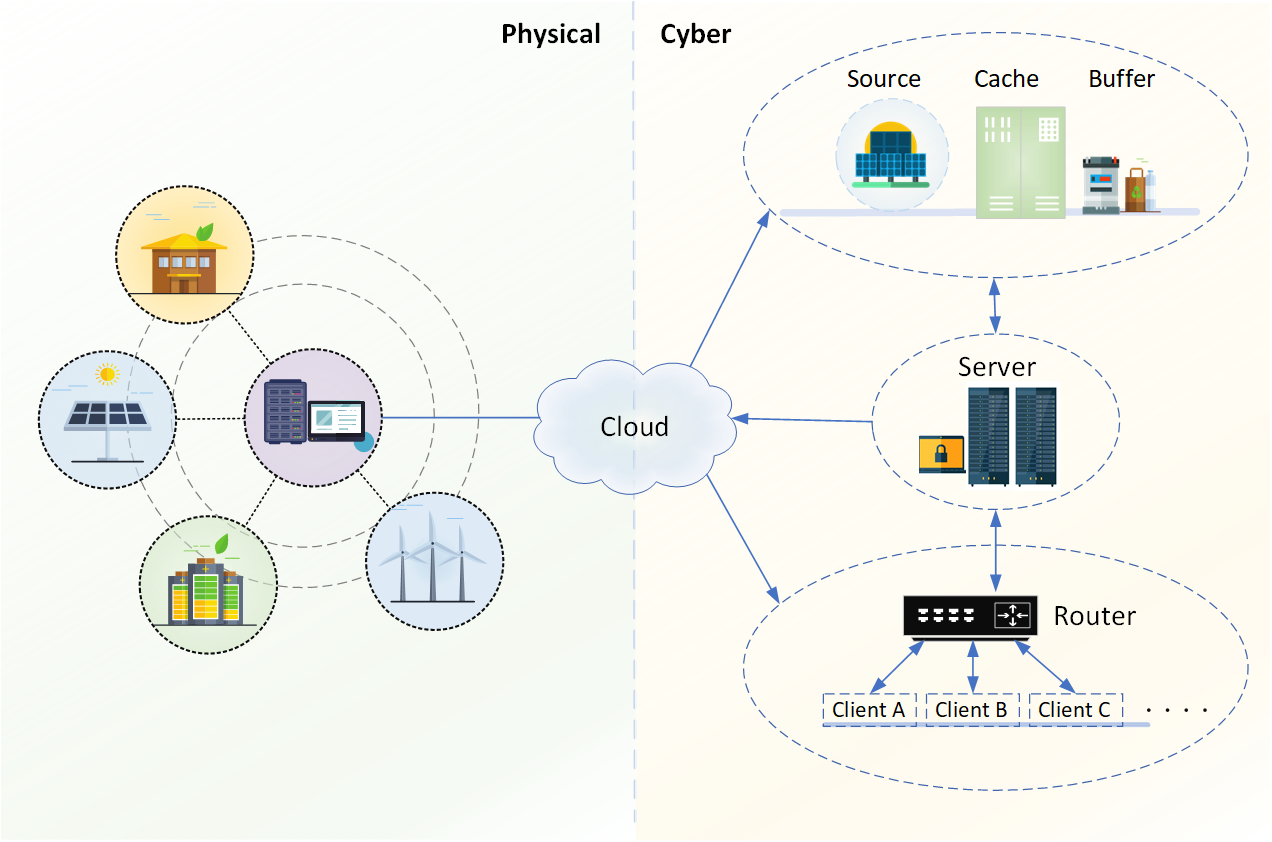}
  \caption{Illustration of the proposed concept. The electricity network is conceptualized as a cyber-physical system based on physical elements that are defined as software-defined agents in the cloud (cyber domain).}
  \label{fig:concept}
\end{figure}

In this study, we focus on DSM for residential loads considering operational decisions in the tertiary-control regime (e.g., 10-min timescale with a limited time horizon of 24 h).
Hence, our solution is similar to the role played by retailers today in European electricity markets, as well as some energy service providers and aggregators.
The proposed SDEN denotes a cyber-physical system employed to manage the (physical) energy delivery as a virtualized energy inventory problem based on \textit{energy packets}, determined in the cyber domain.
Fig. \ref{fig:concept} illustrates the proposed solution.
It is worth noting that this architecture presumes a reliable, secure, and scalable Internet of Things (IoT) network as the key enabler.
This is reasonable and feasible, given the recent rapid developments in the IoT technology.
In addition to the existing wired and dedicated wireless solutions, specialists expect that the fifth generation of mobile system (5G)---designed to manage interconnected physical systems---will become the core of the IoT networks of the future \cite{hui20205g}.

\subsection{Packetized energy management (PEM)} 
\begin{figure}[t]
\centering
  \includegraphics[width=\columnwidth]{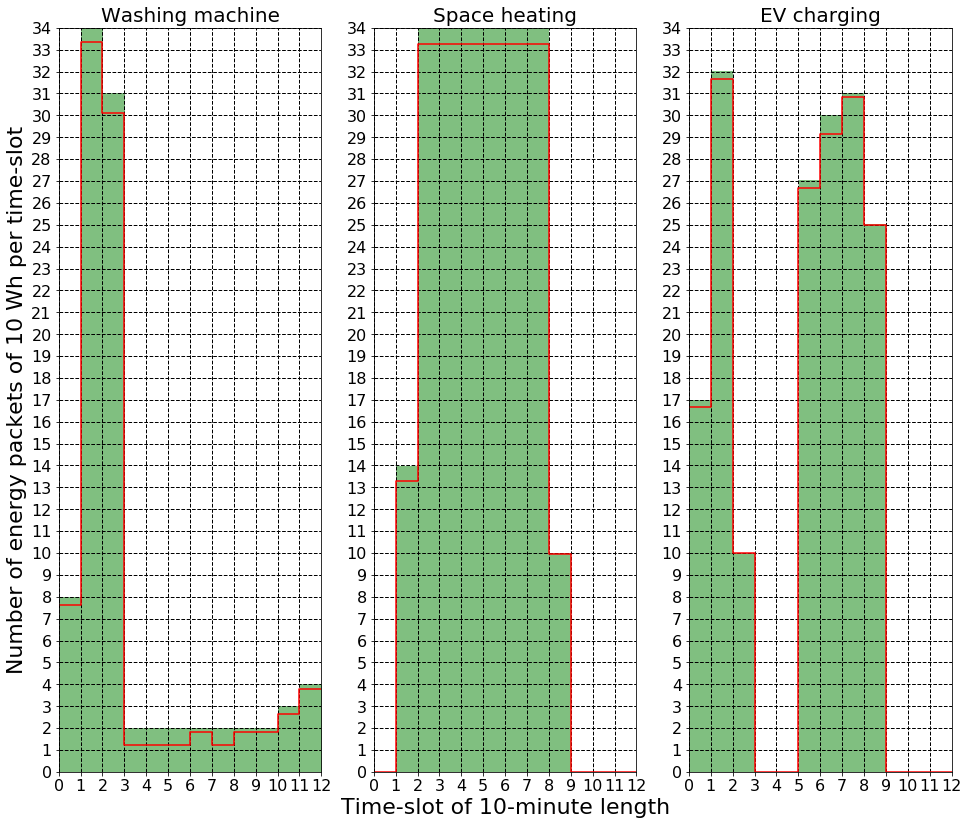}
  \caption{Examples of how loads---here, (a) washing machine, (b) space heating, and (c) EVs---can be packetized.
  }
  \label{fig:pem}
\end{figure}

Among the different ways of approaching DSM, packetized energy management (PEM) as defined in \cite{nardelli2019energy,de2020cyber} provides several advantages due to its capability of multiplexing and prioritization with potentially variable granularity. 
Fig. \ref{fig:pem} shows how PEM maps different flexible loads and storage units into discrete signals composed of energy packets that are defined as chunks of $x$ Wh with a fixed length (i.e., duration) of $y$ min.
In this example, we consider packets of 10 Wh and length of 10 min with three different loads \cite{richardson2010domestic,de2020cyber}: (a) washing machine, (b) space heating, and (c) electric vehicles (EVs).
Note that each load is associated with a different type of flexibility in accordance with their function in the household: (a) the washing machine service needs to be ready at some point of time, but once the machine is on, it cannot be turned off; (b) the space heater must maintain the temperature of the house in a given range all the time, but short-term variations due to thermal inertia, as well as the possibility of operating out of the specified range if there is no one at home, can provide demand flexibility; and (c) the EV must be charged and ready to use when needed, but how the charging is distributed over time is irrelevant.
%
%

In the cyber domain, the flexible loads and storage units are tagged with specific identification details (IDs) that define, for example, their individual service requirements and their priority levels.
Note that DERs related to generation, such as solar also need to be discretized with the same packet sizes.
%
It is also worth mentioning that the cyber-physical PEM considered here is a hybrid of the device-focused PEM solution based on randomization as in \cite{espinosa2020packetized} and the hierarchical physical PEM as in \cite{ma2019rudiment}. 

\subsection{Software-defined energy networks} Virtualization of physical processes has become commonplace.
New concepts like cyber-physical systems and digital twins indicate the interwoven dynamics between material, informational, and decision-making processes \cite{nardelli2018smart}.
Typically, cyber-physical systems are based on relatively autonomous (but interdependent) entities that can act within the system boundaries.
These entities are referred to as \textit{agents} and are often defined via a dedicated software platform \cite{karnouskos2019key}. 
The SDEN proposed in this article is such a platform where cyber (software-defined) agents are designed in accordance with their specific physical \textit{functions}.
The cyber-physical relationships among the agents are determined by PEM that enables the agents to produce, consume, allocate, and store energy packets.
Table \ref{table_agents} presents the nomenclature employed in this study; it is clearly inspired by the conventional naming of computer network elements and Internet peer-to-peer applications \cite{james20016computer}, as well as wireless communication resource allocation methods \cite{popovski2020wireless}.
The agents as well as their inter-related operations are explained in detail in Section \ref{sec:virtual_mrid_as_cps}.
\begin{table*}[t]
\centering
\caption{Agents of the proposed software-defined energy network}
\label{table_agents}
\begin{tabular}{|l|l|l|}
 \hline
\textbf{Cyber element}&\textbf{Physical entity} &\textbf{Examples} \\
 \hline
  \hline
Client    & Flexible load  & EVs, dischwasher, sauna, space heating \\ \hline
Buffer  & Short-term storage & Batteries, heat pumps\\\hline
Cache   & Long-term storage & Synthetic fuel \\ \hline
Source & Aggregated production mainly from distributed generation & Solar PV forming a Virtual Power Plant \\\hline
Router  & Computer or embedded system & Smart meter or home energy management system\\\hline
Server & Computer(s) orchestrating the SDEN & Software in cloud server \\\hline

\end{tabular}
\end{table*}

To coordinate their actions, the SDEN agents need to communicate to either request a service or reply to a service request, as well as schedule actions related to energy delivery or storage.
This communication is defined in the cyber domain based on the following generic types of messages:
\begin{itemize}
    \item \textbf{Request:} Clients send a message to their respective energy server detailing the service they require, which includes, for example, how many energy packets they need, how these packets need to be allocated over time, and the deadline for the service to be accomplished.
    \item \textbf{Acceptance/rejection:} After receiving a request, the server assesses the current and future states of the energy inventory to make a decision, which is then communicated to the respective client with a \textit{yes} or \textit{no} message. In the case of a \textit{yes} message, the client will wait for the delivery/schedule messages. In the  of a \textit{no} message, the server sends a message that will enable the client to plan its new request.
    \item \textbf{Delivery/storage:} The energy server dynamically updates the schedules of energy packets' delivery and storage based on the accepted services and production forecasts. Following the plan, the server sends messages to the clients or storage units indicating that they need to be ready to receive their packets.
    \item \textbf{Acknowledgment:} Clients and servers may also send acknowledgment messages about the delivery of energy packets, in a sense similar to transmission control protocol/Internet protocol (TCP/IP).
    \item \textbf{Emergency:} Some clients may also have the chance of sending messages requesting urgency in the service. For example, change of plans related to EV usage may require urgent messages to be sent to request charging. Such messages should be rare.
\end{itemize}

\begin{figure}[t]
\centering
  \includegraphics[width=\columnwidth]{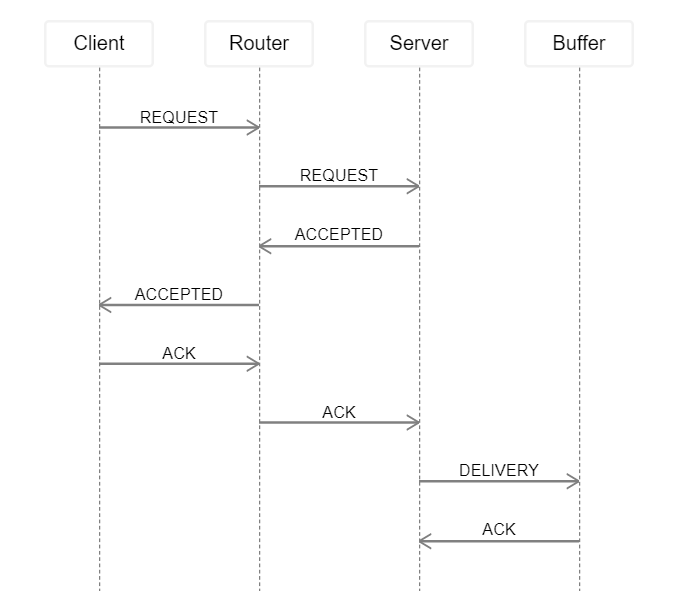}
  \caption{Handshake example: A client requests a service that is accepted by the server and then delivered using the energy available in a given buffer.}
  \label{fig:hand}
\end{figure}

Fig. \ref{fig:hand} illustrates a typical message exchange under the proposed PEM protocol.
This communication between the software agents ultimately determines the cyber-physical system dynamics.
In the following sections, we will present the details of how this generic protocol can be used to build the proposed virtual microgrid and, more importantly, achieve efficient management of microgrid resources.

\section{Virtual Microgrid as a Cyber-physical System}\label{sec:virtual_mrid_as_cps}
\begin{figure}[t]
\centering
  \includegraphics[width=\linewidth]{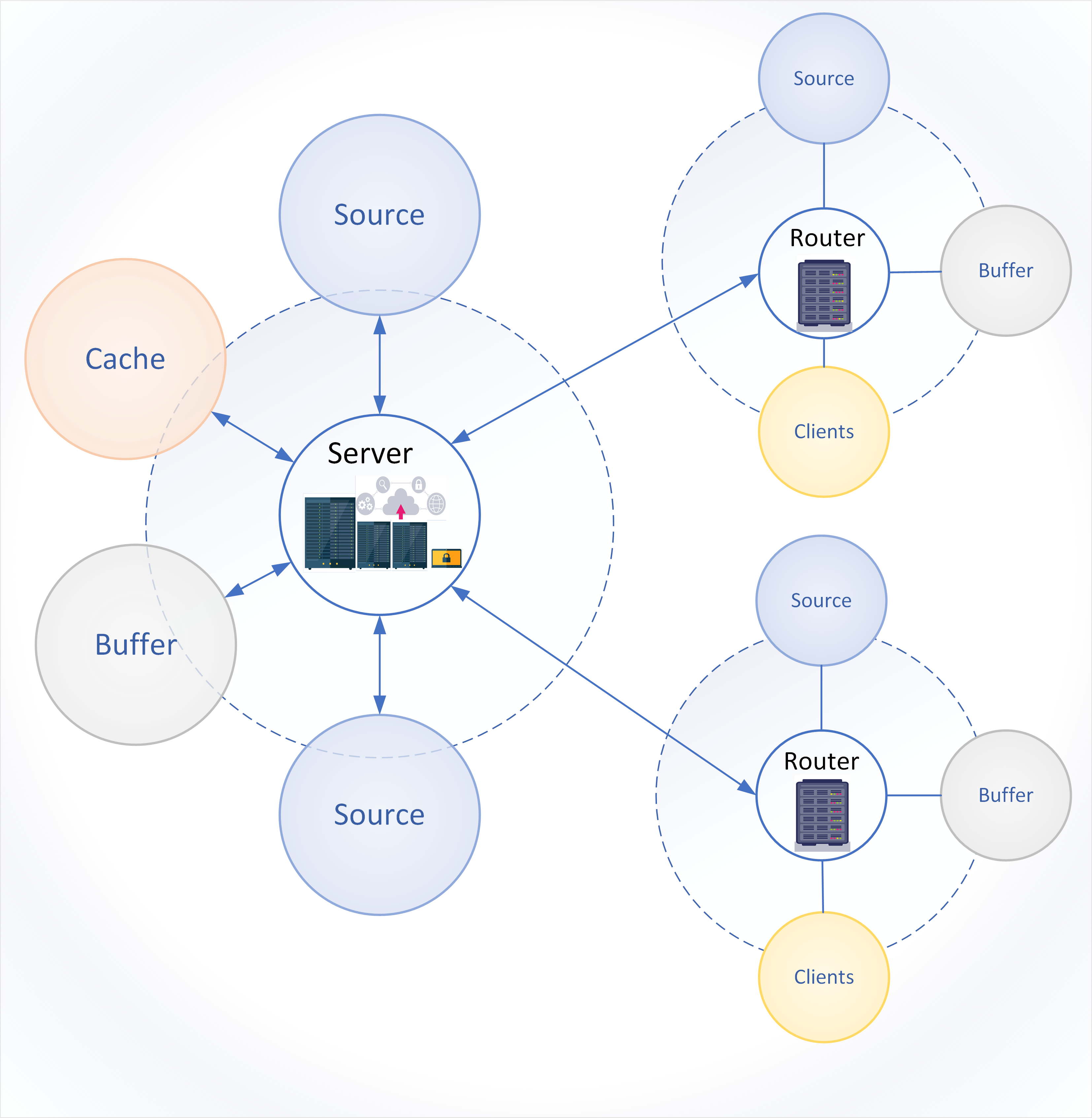}
  \caption{Example of how the agents are related and orchestrated by the Energy Server to build a Virtual Microgrid based on the SDEN elements and communication protocol.}
  \label{fig:virtual-mg}
\end{figure}

The virtual microgrid is constructed with the cyber-physical entities presented in Table \ref{table_agents} and communication among the entities.
For proper operations, the SDEN agents need to be orchestrated following a protocol based on the aforementioned five classes of messages and the prescribed functions of each agent.
Fig. \ref{fig:virtual-mg} illustrates how the proposed virtual microgrid works as a cyber-physical system.
The dynamical behavior of the cyber-physical system can be explained based on the agents' functions, as follows.
\begin{itemize}
    \item \textbf{Clients} are the different flexible loads that send their individual requests to local servers or routers that either locally make decisions about the requests or forward them to the main server. The requests differ depending on the load and the specific kind of flexibility that they can provide.
    \item \textbf{Routers} process and/or aggregate requests from the same household before they are forwarded to the main server. They are either smart meters or home energy management systems. Depending on how the SDEN is orchestrated, they can act as \textit{local servers} to reallocate the local energy packets at that specific house without sending the request to the main server. Otherwise, they forward the request to the main server, acting as routers.
    \item \textbf{Server} processes requests from different routers to make decisions based on the energy available in the shared pool of virtualized energy resources. The packets can be accepted or rejected. The accepted packets need to be scheduled and delivered (this will be discussed later). The rejected requests are returned to the clients who are informed about the decision as well as the procedure for resubmitting a new request. This is the \textit{key agent} in the proposed architecture.
    \item \textbf{Source} refers to the energy produced by distributed sources such as solar photovoltaics (PVs) and wind generators. They are also discretized as identifiable packets that are aggregated as one energy source, thereby building a Virtual Power Plant. Energy from other sources might be incorporated in the source.
    \item \textbf{Buffers} are short-term energy storages such as batteries and heat pumps. They are used to match supply and demand along different operational timescales, ranging from primary to tertiary control. For the tertiary control, the server can allocate some packets from the source to buffers during some periods, or it may allocate some packets from the buffer to supply energy during other periods. This allocation may be done by either a local router or the server.
    \item \textbf{Cache} is related to long-term (seasonal) storage, which is not available today for households but may be integrated in the future. This encompasses, for example, power-to-X technologies to store energy as synthetic fuels. In this case, extra production in the summer may be transferred to the winter. 
\end{itemize}

%

The proposed concept is based on the SDEN-enabled virtual microgrid, where requests and supply are subdivided into energy packets similar to BitTorrent Internet applications \cite{cohen2003incentives,bharambe2006analyzing}.
In operational terms, the resource allocation problem to be solved by the energy server is similar to, for example, university admission processes, where applicants are subdivided into priority levels and the open positions are filled accordingly.
The key difference in our approach is that it has a strong dependence on both past allocations and predicted future availability that is variable due to generation variability and storages.
%
%

The energy server has to (i) admit new energy packets, (ii) (dynamically) classify them into priority classes, (iii) define how many packets of each class can be served in different time slots, and finally (iv) allocate the time when they will be delivered.
Each of these stages has specific challenges, as follows:

\textbf{1. Admission:} A reliable forecast of the energy packet generation for a long period (e.g., 12 h) is necessary. The challenge is that the generation plus storage must match demand. If the server admits more packets than those available on the supply side, some of the clients will not be served, which decreases the quality of the service provided. Another important aspect is the possibility of emergency messages. If a client's requests are rejected many times, the client may send an emergency message to convey that its packets need to be treated with high priority. Emergency messages might also be sent if some appliances or devices have to be used urgently.  Different approaches could be utilized as part of the admission control process, for example, queuing theory, machine learning, and probabilistic forecasting of weather and energy.

\textbf{2. Classification:} The accepted energy packets may be classified into priority classes. For computational purposes, it is preferable that there are limited number of priorities.
    %
    However, the number of classes and techniques to perform the classification are open questions. Energy routers can preprocess the requests and classify it at the household level, or the classification can be related to the appliance type. The classification task can then be built with algorithms based on rules-based approaches, hierarchical classifications, or statistical methods. Advances in machine learning and artificial intelligence promise good performance for dynamically classifying the requests.

\begin{figure}[t]
\centering
  \includegraphics[width=1\columnwidth]{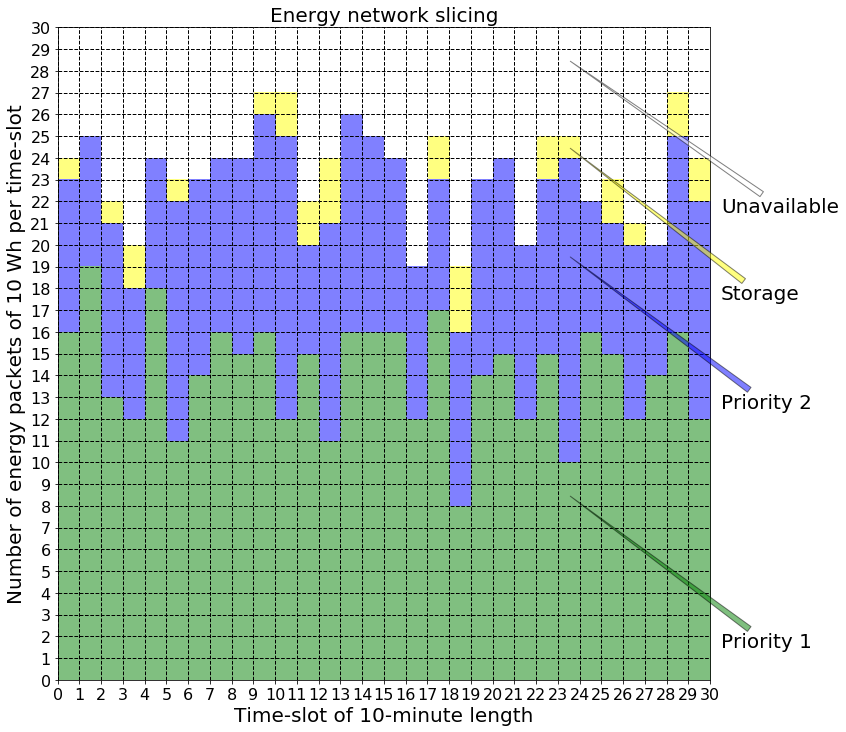}
  \caption{Example of allocation based on network slicing with two priority levels plus additional flexibility available via storage. We consider a time horizon of 30 slots.}
  \label{fig:alloc}
\end{figure}

\textbf{3. Availability:} Once the accepted packets are classified, the server needs to check the availability of the energy packets for upcoming time-slot(s), i.e., how many packets can be served considering the following constraints: (i) the uncontrollable baseload; (ii) the active flexible loads that cannot be turned off after it is started (e.g., dishwashers); (iii) the short-term generation forecasts; and (iv) the possibility of using storage. This will define the system capacity for that specific slot. The capacity can be divided into smaller \textit{slices} based on priority classes, i.e., some packets can be reserved for different packets based on their priority. 
    We call this \textit{energy network slicing} (see Fig. \ref{fig:alloc}). Machine learning and predictive algorithms for inventory management can be expected to provide good supporting tools for achieving this task. 
 
\textbf{4. Allocation:}
    With the packets classified and availability defined, the server needs to select which packets will be allocated for delivery in the upcoming slot(s). In this case, the server will fill the vacant positions of each slice with packets from its associated class. 
    Other prioritization rules may be developed; for example, a closer deadline is first served, or an earlier arrival is first served. Alternative options, such as random rules, can also be devised and used.
    Researches into resource allocation in software-defined networks, including different optimization approaches and queuing theory, may provide good background for building and assessing different approaches to the availability problem.

To illustrate the concept discussed above, let us consider a toy example consisting of a case where the server has two priority levels to classify the packets. Additional flexibility is provided by the energy buffer. The energy packets need to be delivered over the next time slots considering a time horizon of, for instance, 30 slots.
Once the packets are classified, they can be allocated and delivered using \textit{energy network slices} that are simply the number of packets in each priority class that is allocated in a given slot.
It is important to remark that the proposed implementation may be vulnerable to cyber attacks \cite{rawat2016software}.
A new element in the SDEN---mapping the role of the \textit{firewall} in Internet----needs to be designed to guarantee the safe operation of the virtual microgrid.
A systematic investigation of such a protective agent is, however, beyond the scope of this article and should be the focus of future research.

Fig. \ref{fig:alloc} presents the energy network slicing technique considering the following setting: (i) two slices, namely priorities 1 (green) and 2 (blue) considering the available energy from the available sources, and (ii) additional packets available from the buffer/cache (yellow). 
This shows that the proposed management is capable of directly generating a specific demand curve using \textit{packet multiplexing}.
Note that this profile can be dynamically modified to reflect the actual state of the system and to guarantee the balance between supply and demand based on the operational needs of the devices and the wishes of the end users.
Moreover, this approach does not assume any price signals or cost-related optimization techniques and only considers direct usage requests.

\section{Toward a Bright Future with Free Electricity based on Sharing?}
%

The proposed solution aims to achieve a future fossil-fuel-free energy system governed as a commons, i.e., a system with abundant ``clean'' electric energy shared as a commons \cite{giotitsas2020private}.
Such a commons-based renewable electrification paradigm can be achieved by prioritizing electricity sharing within the microgrid over purchasing from the main grid in the proposed solution.
Nevertheless, the solution also needs to comply with the current electricity market structure in order to develop, expand, and eventually become the dominant way that electricity is supplied.
This is the main argument in \cite{nardelli2019energy}, where the authors discussed the concept of the Energy Internet.
In fact, the proposed SDENs and virtual microgrids are not too far from existing commercial solutions (e.g., \cite{klein2019novel}), and so, it is generally compatible with European markets.
Scaling up electricity sharing is also feasible with the use of Exclusive Group (EXG) bidding schemes \cite{kuhnlenz2018implementing,exg}, where different load profile curves are bid in the market.
In the market clearance, a load curve is selected and informed to the bidder, who, in turn, needs to ``generate'' such a demand profile (otherwise, it needs to cover the costs related to up- or down-regulation).
%
%
This approach is an interesting way of managing the flexibility of loads, but it has an important drawback that, if the EXG-type of bids becomes popular in the market, existing problems related to the optimality of the market outcomes (e.g., \cite{transparence-market}) may increase, as discussed in \cite{kuhnlenz2018implementing}. 

%


In summary, we argue that a future with free electricity based on sharing is indeed possible using the technical approach elucidated in this article, although it might not be practical yet under the current mode of production dominated by market relations.
As demonstrated in \cite{bauwens2019peer,giotitsas2019open,kostakis2018convergence}, commons-based peer production is emerging across different sectors and, as pointed out in \cite{giotitsas2020private}, we have all the conditions necessary to make this move in the energy sector as well.

\section{Conclusions}
In this paper, we have proposed a futuristic view of an energy system organized as a cyber-physical system, following a commons-based governance model, whose management is enabled by virtualized energy packets.
In the  proposed cyber-physical  system, electrical  energy sharing  is  enabled  in the cyber domain by handshakes and resource allocation methods similar to those utilized in computer networks. 
To elucidate our proposed approach, we have mapped concepts from modern computer networks and ICTs to the smart grid domain.

One of the main advantages of this approach is that it enables a balance between supply and demand based on direct requests, without overly complex market mechanisms.
Moreover, energy access is also guaranteed to all virtual microgrid  members  according  to  their  individual  needs.
This, however, has the drawback of relying on a home energy management system and wireless connectivity.
We nevertheless do not see this as a strong limitation because cost-effective solutions have been previously developed, for example, in \cite{singh2019smart}.

A more challenging task is to enable the proposed virtual microgrid to handle uncertainty in its operation.
As discussed in Section  \ref{sec:virtual_mrid_as_cps}, all the four stages of the Energy Server require predictions about production and demand, as well as storage capabilities. 
Future research will focus on developing predictive methods as in \cite{kabir2018neural,bassamzadeh2017multiscale}.

\bibliographystyle{IEEEtran}
\bibliography{ref.bib}

\begin{IEEEbiography}[{\includegraphics[width=1in,height=1.25in,clip,keepaspectratio]{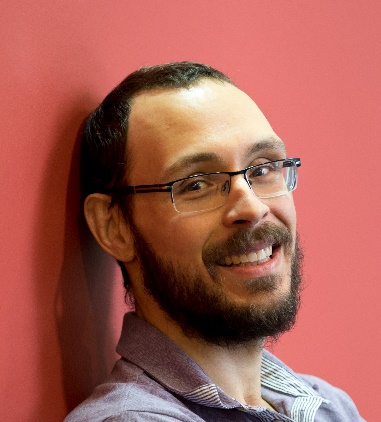}}]{Pedro H. J. Nardelli} received the B.S. and M.Sc. degrees in electrical engineering from the State University of Campinas, Brazil, in 2006 and 2008, respectively. In 2013, he received his doctoral degree from University of Oulu, Finland, and State University of Campinas following a dual degree agreement. He is currently Associate Professor (tenure track) in IoT in Energy Systems at LUT University, Finland, and holds a position of Academy of Finland Research Fellow with a project called Building the Energy Internet as a large-scale IoT-based cyber-physical system that manages the energy inventory of distribution grids as discretized packets via machine-type communications (EnergyNet). He leads the Cyber-Physical Systems Group at LUT and is Project Coordinator of the CHIST-ERA European consortium Framework for the Identification of Rare Events via Machine Learning and IoT Networks (FIREMAN). He is also Docent at University of Oulu in the topic of “communications strategies and information processing in energy systems”. His research focuses on wireless communications particularly applied in industrial automation and energy systems. He received a best paper award of IEEE PES Innovative Smart Grid Technologies Latin America 2019 in the track “Big Data and Internet of Things”. He is also IEEE Senior Member. More information: https://sites.google.com/view/nardelli/
\end{IEEEbiography}

\begin{IEEEbiography}[{\includegraphics[width=1in,height=1.25in,clip,keepaspectratio]{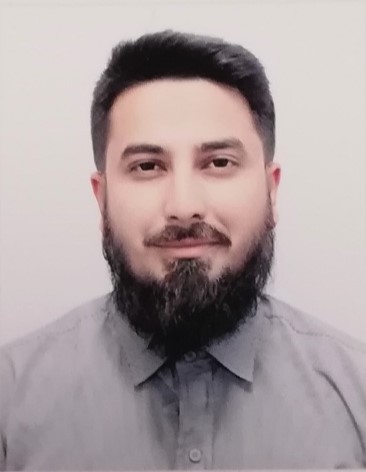}}]{Hafiz Majid Hussain} (M'19) completed his BS and MS in Electrical Engineering from the National University of Computer \& Emerging Sciences and University of Engineering \& Technology Taxila, Pakistan, in 2014 and 2017, respectively. He is the part of the project called Building the Energy Internet as a large-scale IoT-based cyberphysical system. Currently, he is pursuing a Ph.D. towards Electrical Engineering from the Lappeenranta University of Technology in the research group Cyber-Physical Systems Group, Finland. His research interest includes demand response applications, energy resource optimization in smart grid and energy internet, and information security technologies.More information: https://sites.google.com/view/hafizmajidhussain/biography
\end{IEEEbiography}

\begin{IEEEbiography}[{\includegraphics[width=1in,height=1.25in,clip,keepaspectratio]{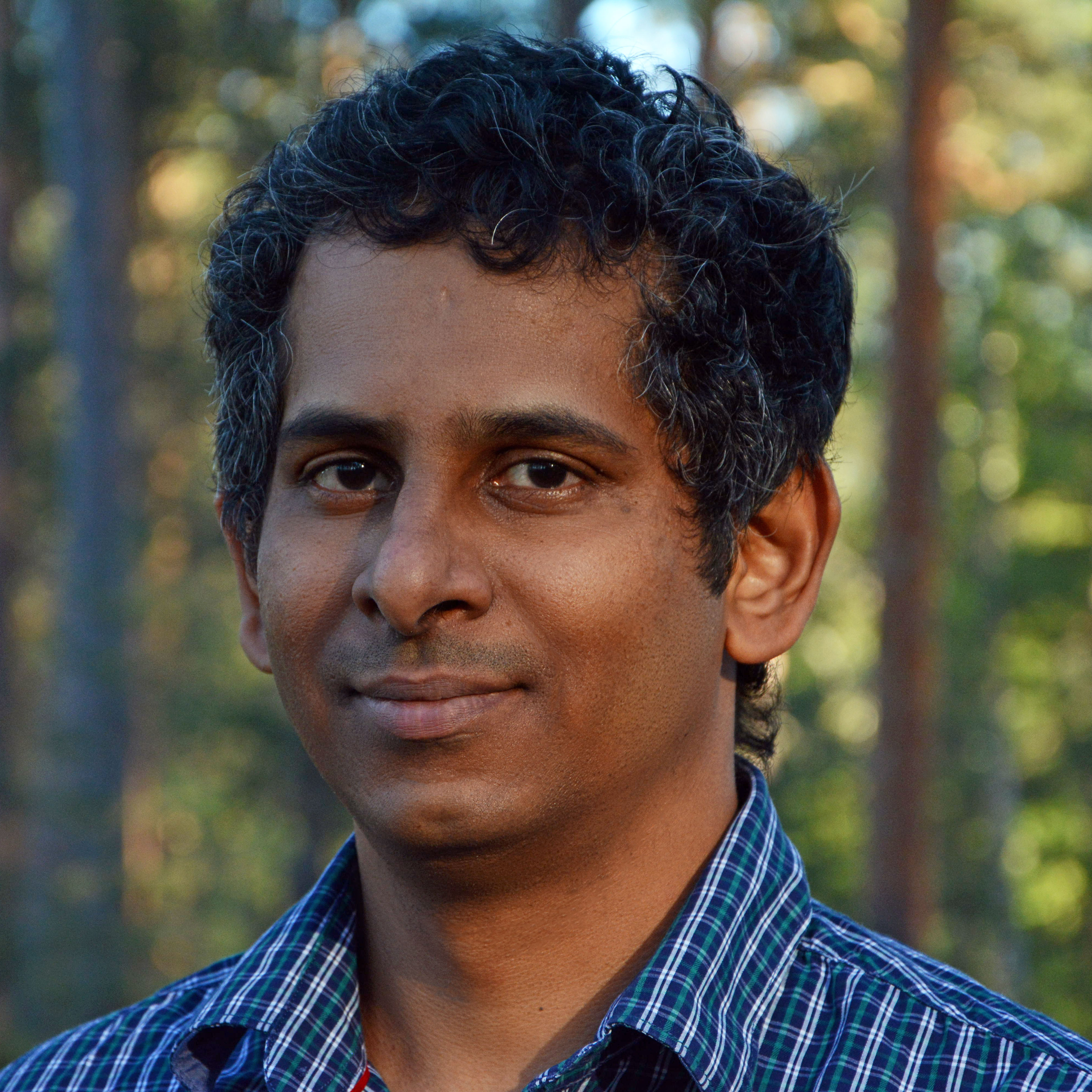}}]{Arun Narayanan} (M'14) received his B.E. degree in Electrical Engineering from Visvesvaraya National Institute of Technology, Nagpur, India and M.Sc. in Energy Technology from Lappeenranta University of Technology (LUT), Finland, in 2002 and 2013, respectively. He subsequently completed his Ph.D. from the School of Energy Systems, LUT University. 
He is currently a Postdoctoral Fellow with LUT University, Lappeenranta, Finland, in the research group Cyber-Physical Systems Group. His research interests include renewable energy-based smart microgrids, electricity markets, demand-side management, energy management systems, and information and communications technology. He focuses on applying optimization, computational concepts, and artificial intelligence techniques to renewable electrical energy problems.
\end{IEEEbiography}

\begin{IEEEbiography}[{\includegraphics[width=1in,height=1.25in,clip,keepaspectratio]{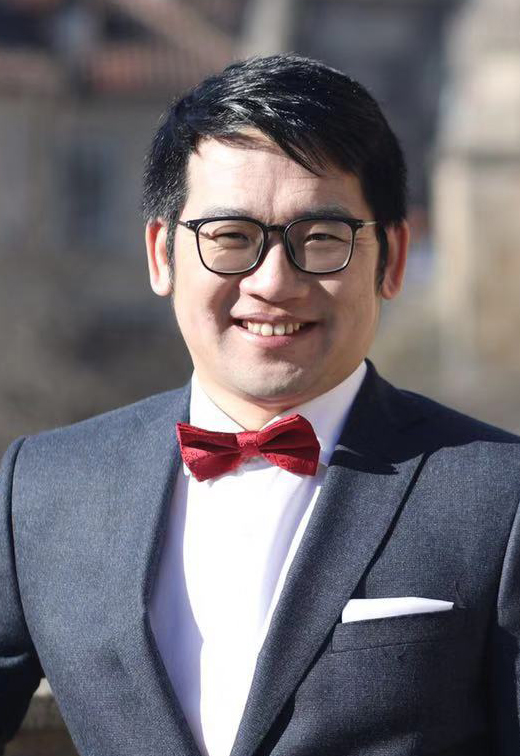}}]{Yongheng Yang} (SM’17) received the B.Eng. degree in electrical engineering and automation from Northwestern Polytechnical University, Shaanxi, China, in 2009 and the Ph.D. degree in electrical engineering from Aalborg University, Aalborg, Denmark, in 2014.
He was a postgraduate student with Southeast University, China, from 2009 to 2011. In 2013, he spent three months as a Visiting Scholar at Texas A{\&}M University, USA. Currently, he is ZJU100 Professor, Zhejiang University, China.
Previously, he was an Associate Professor with the Department of Energy Technology, Aalborg University, where he also served as the Vice Program Leader for the research program on photovoltaic systems. His current research is on the integration of grid-friendly photovoltaic systems with an emphasis on the power electronics converter design, control, and reliability.
Dr. Yang was the Chair of the IEEE Denmark Section. He serves as an Associate Editor for several prestigious journals, including the {\sc IEEE Transactions on Industrial Electronics, the IEEE Transactions on Power Electronics}, and the IEEE Industry Applications Society (IAS) Publications. He is a Subject Editor of the IET Renewable Power Generation for Solar Photovoltaic Systems. He was the recipient of the 2018 IET Renewable Power Generation Premium Award and was an Outstanding Reviewer for the {\sc IEEE Transactions on Power Electronics} in 2018.
\end{IEEEbiography}

\end{document}